\documentclass[%
 reprint,
 amsmath,amssymb,
 aps,
]{revtex4-2}

\usepackage{graphicx}
\usepackage{dcolumn}
\usepackage{bm}
\usepackage{tcolorbox}
\usepackage{float}

\begin{document}

\preprint{APS/123-QED}

\title{Synchronization of complex spatio-temporal dynamics with lasers}
\author{Jules Mercadier$^{*,1,2}$}
\author{Stefan Bittner$^{1,2}$}
\author{Marc Sciamanna$^{*,1,2}$}
\affiliation{%
$^1$Université de Lorraine, CentraleSupélec, LMOPS, Metz, France\\
$^2$Chaire Photonique, LMOPS, CentraleSupélec, Metz, France \\
$^*$ Corresponding authors : jules.mercadier@centralesupelec.fr (+33664193854); marc.sciamanna@centralesupelec.fr (+33632829331)
}%

\begin{abstract}
Synchronization is the spontaneous alignment of the dynamics of weakly-coupled oscillators. In addition to temporal dynamics like periodic and chaotic oscillations, also the spatio-temporal dynamics of spatially-extended systems like wildlife populations can synchronize. We exploit here the intrinsic spatio-temporal complex dynamics of broad area lasers to demonstrate such synchronization at lab-scale. Broad-area vertical-cavity surface-emitting lasers (BA-VCSELs) exhibit chaos from the nonlinear coupling between laser modes with different spatial profiles and polarization. When coupling two BA-VCSELs, several synchronization and anti-synchronization regimes are observed, highlighting the complex interplay between oscillating modes with different frequencies and spatial patterns. The correlation coefficient varies between 0.2 and 0.9 depending on the dynamics and on the time scale under analysis. Besides its fundamental interest, our experiment with commercial devices marks the first step towards real-world spatial multiplexing in multiple user physical-layer secure communication based on chaos synchronization. 
\end{abstract}

\maketitle


\section*{Introduction}
Synchronization between coupled oscillators is a phenomenon encountered in a variety of fields including electronics \cite{csaba_coupled_2020,parihar_exploiting_2014}, chemical reactions \cite{forrester_arrays_2015}, biological networks \cite{mirollo_synchronization_1990}, and lasers \cite{hui_injection_1991, lang_injection_1982,mogensen_locking_1985}. In the simplest case of two coupled harmonic oscillators, synchronization means the two oscillators are frequency- and phase-locked, which was first observed for two coupled pendula \cite{bennett_huygenss_2002}, and is usually described by the Adler equation \cite{Adler1946}. Synchronization of clocks is of immense practical importance for time and frequency metrology. 

More generally, synchronization means the ability of weakly coupled systems to develop correlated temporal dynamics. Synchronization can also occur between coupled chaotic systems, in spite of their inherent sensitivity to initial conditions \cite{pecora_synchronization_1990, sciamanna_physics_2015, vanwiggeren_communication_1998, argyris_chaos-based_2005}. Chaos synchronization is typically achieved with the master-slave unidirectional coupling configuration \cite{mirasso_synchronization_1996, rogister_secure_2001}, but also common-signal induced synchronization has been reported. \cite{Yamamoto:07, wang_experimental_2025}. While temporal synchronization has been thoroughly explored, the extension to spatio-temporal systems remains largely unexplored. Systems with significant transverse extent can exhibit complex space- and time-dependent dynamics, but the spatial dimension introduces new challenges for achieving and understanding synchronization. 

Synchronization of complex spatio-temporal dynamics occurs naturally for example in ecological systems \cite{blasius_chaos_2000, blasius_complex_1999}, the human brain \cite{barardi_phase-coherence_2014} and weather patterns \cite{marwan_complex_2015}. Understanding and replicating these dynamics in artificial systems and at lab scale could revolutionize neuroscience, machine learning, biology and our understanding of complex systems in general. 

Optical systems are an ideal testbed for studying nonlinear dynamics thanks to their compact size and fast time scales \cite{sciamanna_physics_2015}. Complex spatio-temporal dynamics like pattern formation, filamentation \cite{adachihara_spatiotemporal_1993, fischer_complex_1996}, and modal competition \cite{lenstra_spatio-temporal_2004} can be found for example in broad-area semiconductor lasers. In addition, theoretical studies have predicted the possibility to synchronize multimode and distributed optical systems~\cite{buldu_multimode_2004, garcia-ojalvo_spatiotemporal_2001} and shown their potential for parallel chaotic information processing and multiple user encryption schemes~\cite{white_multichannel_1999, banerjee_synchronization_2011}. Nonetheless synchronization of complex spatio-temporal dynamics has been experimentally demonstrated so far only using a liquid crystal light valve with feedback \cite{havermann_synchronisation_2008}, and its implementation therefore remains far from practical physical-layer communication architectures. 

We investigate the possibility to synchronize spatio-temporal chaotic systems using commercially available lasers in a table-top experiment. We take advantage of our recent discovery \cite{bittner_complex_2022, mercadier_chaos_2025} that broad area vertical cavity surface-emitting lasers (BA-VCSELs) naturally exhibit chaos involving nonlinear interactions between a large number of spatial modes with different polarization states \cite{valle_dynamics_1995, barchanski_picosecond_2003, GIUDICI1998313, Buccafusca1996, buccafusca_transient_1999}. By coupling two such lasers, we demonstrate synchronization in a controllable and reproducible way and analyze how synchronization relates to the spatial lasing modes. Besides its fundamental interest for the study of complex systems, our work marks the first step to a realistic implementation of spatial multiplexing for physical-layer secure communications at high bit rates.

\begin{figure*}
    \centering
    \includegraphics[width=\linewidth]{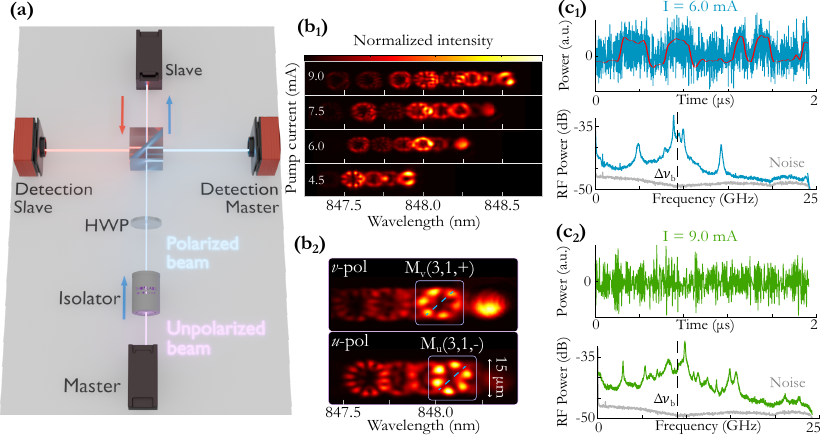}
    \caption{Setup and free-running master laser. (a) Schematic representation of the synchronization experiment setup. This configuration allows both the analysis of the dynamics of the master laser, and the response of the slave laser under injection. It is possible to study both of the lasers in the spatio-spectral domain, using an imaging spectrometer (with a half-wave plate (HWP) to control the polarization orientation at the spectrometer input), or in the temporal domain, using two high-bandwidth photodetectors connected to a high-bandwidth real-time oscilloscope. (b$_1$) Spatio-spectral images of the free-running master laser for different pump currents in $u$-polarization, which is selected for injection; (b$_2$) Spatio-spectral image in both $u$- and $v$-polarizations for $I = 6.5$~mA, with the strong modes $M_{u,v}(3, 1)$ highlighted. The symmetry axis of the transverse mode is given by the blue dashed line to indicate the + (-) orientation of the modes. Master laser dynamics for $I_M$= 6 mA (c$_1$) and $I_M$= 9 mA (c$_2$) in the time and frequency domains. The vertical dashed lines in the RF-spectra indicate the birefringence $\Delta \nu_b$. The red time trace in (c$_1$) is low-pass filtered at $0.1$~GHz.}
    \label{Setup_free_running_V5}
\end{figure*}

Specifically, we investigate two coupled BA-VCSELs which exhibit chaotic dynamics in free-running operation (see Fig.~\ref{Setup_free_running_V5}). Via the pump current, we can change the number of transverse lasing modes and the dynamical regimes~\cite{bittner_complex_2022, mercadier_chaos_2025}. Synchronization of chaotic dynamics is demonstrated for weak coupling via unidirectional optical injection. Remarkably, synchronization emerges despite significant differences in the spatial mode structures. It is mediated by the spectral alignment of the strong transverse modes of the master laser, defined here as those containing a significant portion of its total power. This reveals a mechanism in which frequency alignment prevails over spatial matching in enabling chaos synchronization in multimode systems. Spatial complexity, instead of preventing synchronization, therefore coexists with robust collective dynamics under appropriate coupling conditions. 

We propose to investigate the effect of unidirectional optical injection between two BA-VCSELs in two distinct dynamical regimes: a first regime involving slow polarization-hopping dynamics, and a second regime featuring broadband chaotic dynamics.

\section*{Results}

We experimentally study unidirectional optical injection of one BA-VCSEL into another BA-VCSEL of the same model and analyze the dynamical response of the slave laser to the signal of the master laser. A simplified schematic of the experimental setup is shown in Fig.~\ref{Setup_free_running_V5}(a), and a detailed description is provided in Supplementary Section~\ref{sec:setup}. We begin with a summary of the properties of the master laser (see also Refs.~\cite{bittner_complex_2022, mercadier_chaos_2025}), and a comparison to the slave laser is presented in Supplementary Section \ref{sec:free-running}. 

The free-running BA-VCSELs lase in several transverse modes and two linear orthogonal polarization states, denoted by $u$ and $v$. The dominant polarization depends on the pump current and changes at several polarization-switching points (PSPs, see Supplementary Fig.~\ref{fig:SM_LI_Curves}) \cite{bittner_complex_2022}. The birefringence splitting between $u$- and $v$- polarization modes is around $\Delta \nu_b \approx 9$~GHz (see Supplementary Section~\ref{sec:free-running}), and we refer to "red" ($u$) and "blue" ($v$) polarization according to their relative position in the optical spectrum. 
	
Figure~\ref{Setup_free_running_V5}(b) shows spatio-spectral images measured with an imaging spectrometer (see Supplementary Section~\ref{sec:setup}) for different pump currents. With increasing pump current, more transverse modes are progressively excited. A transverse mode with mode indices $(m, n)$ has $2m$ intensity maxima in azimuthal direction and $n$ maxima in radial direction. Furthermore, for $m>0$, two distinct spatial orientations are possible, which we denote with $+$ ($-$) when they are (anti-) symmetric with respect to a symmetry axis as shown in Fig.~\ref{Setup_free_running_V5}(b$_2$). In the following we denote the modes with $X(m, n, \pm)$ where $X \in \{M, S\}$ refers to the master and slave laser, respectively. It should be noted that modes $(m, n, +)$ and $(m, n, -)$ are not necessarily degenerate \cite{Lin2017}, but can exhibit a splitting $\Delta \nu_O$ of the order of several GHz (see Supplementary Section~\ref{ssec:orientation-splitting}). The actual splitting seems to depend on the transverse mode. Typically, only one of the orientations is excited for a given polarization, and the corresponding mode in the other polarization has the opposite orientation due to gain competition \cite{Debernardi2002, MartinRegalado1997, MartinRegalado1997c, bittner_complex_2022}. This is exemplified in Fig.~\ref{Setup_free_running_V5}(b$_2$) with the modes $M_v(3, 1, +)$ and $M_u(3, 1, -)$. However, in some cases both orientations of a transverse mode $(m, n)$ lase in the same polarizations as shown in Supplementary Fig. \ref{fig:SM_splitting_transverse}. 

The competition of lasing modes with different spatial profiles and polarizations results in rich nonlinear dynamics, as shown in Fig.~\ref{Setup_free_running_V5}(c). As the pump current increases, the system undergoes a sequence of bifurcations that progressively enrich the temporal behavior \cite{bittner_complex_2022, mercadier_chaos_2025}: initially periodic oscillations transition to quasi-periodic regimes, and eventually to chaotic dynamics (see Supplementary Fig.~\ref{fig:SM_RF_Spectra}). These bifurcations are usually accompanied by a redistribution of power between different transverse modes and in some cases polarization switching points \cite{bittner_complex_2022}. We restrict our analysis to parameter regions in which the master laser features chaotic dynamics for a focused investigation of synchronization in the presence of intrinsic multimode chaos. 

The lasing dynamics features different time scales: the dominant frequencies are around the birefringence $\Delta \nu_b$ [see Fig.~\ref{Setup_free_running_V5}(c)]. However, in some current regimes like around $6$~mA, this fast dynamics coexists with a slower polarization-hopping dynamics with frequencies of the order of $100$~MHz [see Fig.~\ref{Setup_free_running_V5}(c$_1$) and Supplementary Fig.~\ref{fig:SM_hopping}], similar to observations for single-mode VCSELs \cite{martin-regalado_polarization_1997, willemsen_polarization_1999, virte_deterministic_2013, olejniczak_polarization_2011}. 

Next, we implement optical injection between two BA-VCSELs to investigate synchronization between them. Although the two lasers are of the same model, some differences in their spectral and dynamical properties inevitably remain. The comparison of their optical spectra and dynamics in Supplementary Section~\ref{ssec:laser-comp} shows that spectral alignment can be achieved by adjusting the detuning via the temperature or the pump current of the slave (see Supplementary Section~\ref{ssec:param-depend}). The two distinct dynamical regimes in Fig.~\ref{Setup_free_running_V5}(c) are injected into the slave laser in order to analyze the synchronization quality for both high frequency chaos and slower mode-hopping dynamics. Different synchronization mechanisms emerge depending on the properties of the injected dynamics. The response of the slave laser strongly depends on the temporal and spectral structure of the injected signal, highlighting the richness of coupling phenomena in chaotic multimode systems.


\begin{figure*}
    \includegraphics[width=\linewidth]{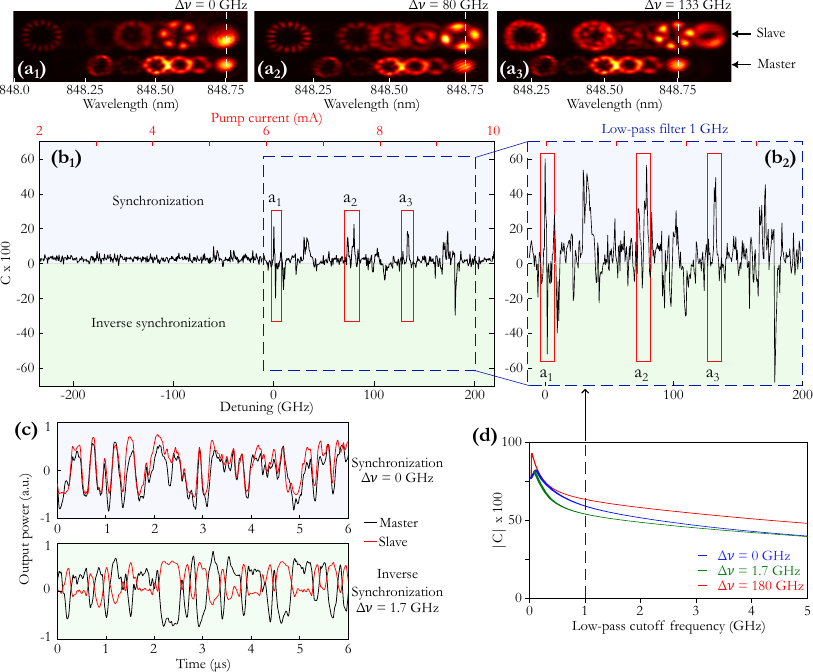} 
    \caption{Synchronization of polarization-hopping dynamics. Spatio-spectral images in $u$-polarization of master and slave laser for $\Delta \nu =$ 0 GHz (a$_1$), 80 GHz (a$_2$), and 133 GHz (a$_3$). The camera field of view corresponds to 2600 $\mu m$ $\times$ 740 $\mu m$ in the object plane. (b$_1$) Correlation between the unfiltered and (b$_2$) low-pass filtered (cutoff at 1~GHz) time traces in $u-$polarization of master and slave laser as a function of the detuning or pump current $I_S$. (c) Low-pass filtered time traces (cutoff at 0.1~GHz) of the master (black) and slave (red) lasers for $\Delta \nu = 0$~GHz (top) and $\Delta \nu = 1.7$~GHz (bottom), corresponding to a correlation of approximately $\pm$80\%, respectively. (d) Absolute value of correlation for  $\Delta \nu = 0$ GHz (blue), $\Delta \nu = 1.7$ GHz (green), and $\Delta \nu = 180$ GHz (red) as a function of the low-pass filter cutoff frequency. The vertical blue dashed lines indicate the cutoff value chosen to compute the correlation coefficient shown in (b$_2$).} 
    \label{correlation-6mA} 
\end{figure*}

The first case we study is when the master laser operates at $I_M = 6.17$~mA and exhibits a dynamics characterized by both high-frequency components and slow polarization-hopping (typically around $100$~MHz). The master laser current and temperature are fixed to keep its dynamical properties constant. The slave laser current is varied between $I_S = 2$~mA and $10$~mA at a fixed temperature of $20^\circ$C. The slave laser modes are red-shifted due to Joule heating as its current increases [see Supplementary Section~\ref{ssec:param-depend}]. Furthermore, its output power increases, thereby reducing the optical injection ratio (see Supplementary Section~\ref{ssec:injection-ratio}). We study parallel injection, meaning that the $u$-polarization component of the master laser, which is its dominant polarization for this pump current, is injected into the $u$-polarization of the slave laser using a HWP. The slave laser has several PSPs in this current range, but its red polarization ($u$) remains mostly dominant (see Supplementary Fig.~\ref{fig:SM_LI_Curves}). 

Figure~\ref{correlation-6mA}(b$_1$) shows the correlation between the time traces (see Methods) of $u$-polarized emissions of master and slave laser as function of the frequency detuning $\Delta \nu$ between the two lasers, which is defined here as the frequency of mode $M_u(0, 1)$ minus the frequency of $S_u(0, 1)$ as illustrated in Fig.~\ref{correlation-6mA}(a$_1$). We observe several cases of significant positive and negative values of correlation, corresponding to synchronization and inverse synchronization respectively, which are discussed in the following. 

The positive correlation peaks are related to spectral alignment between transverse modes of the master and the slave laser with high power. For $\Delta \nu = 0$, the spatio-spectral image in Fig.~\ref{correlation-6mA}(a$_1$) shows that both the $S_u(0, 1)$ and the $S_u(3, 1)$ modes emit strongly and are spectrally aligned with the $M_u(0, 1)$ and $M_u(3, 1)$ modes, respectively. We originally expected this to yield the best synchronization since theoretically each transverse mode of the master laser would be aligned with its counterpart of the slave, but in practice perfect matching of all transverse mode frequencies is not achievable due to small differences in modal spacings between the two lasers. We also observe that for $\Delta \nu = 0$, when the $S_u(3, 1)$ mode aligns with the $M_u(3, 1)$ mode, its power significantly increases while neighboring modes become weaker, as discussed in Supplementary Fig. \ref{fig:SM_Slave_NF_Excited}. This highlights the strong interaction of these two modes as their frequencies align. 

Similar observations are made for the other positive correlation peaks. At $\Delta \nu \approx 30$~GHz, a weak synchronization peak appears, associated with the alignment of $M_u(0, 1)$ with $S_u(1, 1)$. The peak at $\Delta \nu \approx 80$~GHz corresponds to the alignment between $M_u(0,1)$ and $S_u(2,1)$ [see Fig.~\ref{correlation-6mA}(a$_2$) and Supplementary Fig.~\ref{fig:SM_Slave_NF_Excited}]. Similarly, at $\Delta \nu \approx 133$~GHz, another peak emerges from the alignment between modes $M_u(0,1)$ and $S_u(3,1)$ [see Fig.~\ref{correlation-6mA}(a$_2$)]. These examples demonstrate that synchronization can be achieved when a strong transverse mode of the master laser spectrally aligns with a transverse mode of the slave, however, it need not be the same transverse modes: spectral alignment appears to be more important than spatial alignment for successful synchronization. It should also be emphasized that the other transverse modes of the slave laser remain active and are different from those of the master laser [see Fig.~\ref{correlation-6mA}(a)], meaning that synchronizing the temporal dynamics of the lasers does not require synchronizing the optical spectrum or the spatial intensity distribution. 

We also observe inverse synchronization, that is negative correlation between the time traces of the $u$-polarized emission of master and slave, with the most significant examples at $\Delta \nu = 1.7$~GHz and at $\Delta \nu = 180$~GHz. Synchronization in antiphase for coupled oscillators has been known since early studies of coupled pendula and has also been observed in coupled lasers with feedback, for example in the low-frequency fluctuation regime \cite{sivaprakasam_inverse_2001}. Polarization dynamics was identified as a key factor in the emergence of inverse synchronization in Refs.~\cite{hong_synchronization_2004, wedekind_synchronization_2002}. 

The first example at $\Delta \nu = 1.7$~GHz happens right after the alignment of the $M_u(3, 1)$ mode with the $S_u(3, 1)$ mode when the former approaches the frequency of the $S_v(3, 1)$ mode, as shown by the time traces of the master and slave, low-pass filtered at 0.1 GHz cutoff, in Fig.~\ref{correlation-6mA}(c). Indeed, Supplementary Fig.~\ref{fig:spatio-spec-inv-synch}(a) shows that the $S_v(3, 1)$ mode is strongly excited at $\Delta \nu = 1.7$~GHz. We surmise that when injecting light into the $u$-polarization at a frequency near a $v$-polarized mode of the slave, it can create a synchronization of the $v$-polarized emission with the mode-hopping dynamics of the $u$-polarized injection signal from the master, though this is evidently not always the case. Since the emission in $u$- and $v$-polarization is highly anticorrelated in the polarization-hopping regime [see Supplementary Fig.~\ref{fig:SM_hopping}], the $u$-polarized emission of the slave becomes anti-correlated to the $u$-polarized master signal since the latter is synchronized with the $v$-polarized emission of the slave. At $180$~GHz detuning, we observe that the $S_v(1, 2)$ mode is enhanced when its frequency comes close to the $M_v(0, 1)$ mode.

These examples demonstrate that coupling of two multimode VCSELs can create different types of synchronization depending on system parameters, and that moreover strong dynamic correlations can be established between transverse modes of very different order. 

Finally we analyze the master-slave correlation across different timescales by applying spectral filtering (see Methods). Figure~\ref{correlation-6mA}(d) shows the evolution of the correlation as function of the low-pass filter cutoff frequency for three examples. The correlation significantly improves with correlations of up to 90\% for $100$~MHz cutoff at $\Delta \nu$ = 180 GHz. Furthermore, Fig.~\ref{correlation-6mA}(b$_2$) shows the correlation of the low-pass filtered time traces with 1~GHz cutoff as a function of detuning. A global increase is observed across all correlation regions. These low-frequency components represent the relatively slow polarization-hopping dynamics which appears to synchronize much better than the high-frequency components of the dynamics as demonstrated by the low-pass filtered time traces of master and slave laser at $\Delta \nu = 0$ and $1.7$~GHz in Fig.~\ref{correlation-6mA}(c). This demonstrates that it is the polarization-hopping dynamics which is synchronized, and that one can achieve very high synchronization quality even though the BA-VCSELs are not perfectly identical. 

\begin{figure} 
    \includegraphics[width=\linewidth]{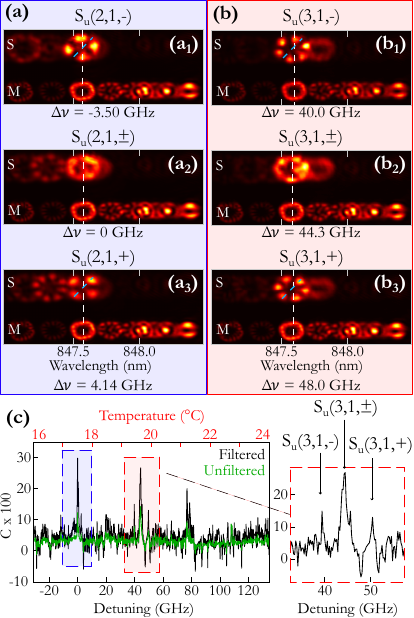} 
    \caption{Synchronization of broadband chaotic dynamics. (a,b) Spatio-spectral images of $u$-polarized emission of the slave laser under injection (top) and the master laser (bottom) for $I_M = 8.8\ \mathrm{mA}$ and $I_S = 4.8\ \mathrm{mA}$. The camera field of view corresponds to 2470 $\mu m$ $\times$ 950 $\mu m$ in the object plan. The detunings are $\Delta \nu =$ -3.50 GHz (a$_1$), 0 GHz (a$_2$), 4.14 GHz (a$_3$), 40 GHz (b$_1$), 44.3 GHz (b$_2$), and 48.0 GHz (b$_3$). The symmetry axis of the transverse modes is indicated with the blue dashed lines. (c) Correlation between the $u-$polarized time traces of master and slave laser as a function of the detuning, without filter (green) and with low-pass filter at 1 GHz cutoff (black). The magnification on the right shows the three-peak scenario around $\Delta \nu =$ 44.3 GHz.} 
    \label{correlation-8.8mA} 
\end{figure}

For the second case of injection that we study, the master laser is set to $I_M=8.8$ mA and $T_M = 20^\circ$C, the slave laser is operated at a fixed current of $I_S = 4.8$~mA, and the detuning is varied via the temperature of the slave. As before, the master laser emits predominantly along the $u$-polarization, and parallel injection of the $u$-polarization is performed. These experimental parameters lead to three differences to the first case. First, the injection ratio is higher though it remains relatively weak (see Supplementary Section~\ref{ssec:injection-ratio}). Second, while the master laser has higher power, it is distributed over a larger number of transverse modes [see Fig.~\ref{correlation-8.8mA}(a)]. Third, the master laser operates in a chaotic state with high frequency components (see Fig.~\ref{Setup_free_running_V5}(c) and Ref.~\cite{mercadier_chaos_2025}) without polarization-hopping. In the following we discuss how these changes affect the synchronization. 

The detuning $\Delta \nu$ is controlled by the temperature of the slave laser, with a measured tuning coefficient of $\Delta \nu / \Delta T = 20.7~\mathrm{GHz/K}$. We define here that $\Delta \nu = 0$~GHz when the mode $S_u(2, 1)$ is aligned with the mode $M_u(6, 1)$, which is the dominant mode of the master and plays an important role in the following. The spatio-spectral images in Fig.~\ref{correlation-8.8mA} reveal that this mode lases in both orientations, $M_u(6, 1, +)$ and $M_u(6, 1, -)$, at the same time, and with a negligible splitting $\Delta \nu_O$. This property of the master strongly influences the synchronization behavior. 

Figure~\ref{correlation-8.8mA}(c) shows several regions of positive correlation. The region around $\Delta \nu \approx 0$~GHz stems from synchronization of the $M_u(6, 1, \pm)$ modes with the $S_u(2, 1, \pm)$ modes, which are very weak in free-running operation but can be strongly excited by the injection. A closer look shows that there are actually three successive correlation peaks: $S_u(2, 1, -)$ is excited at $\Delta \nu = -3.5$~GHz, both $S_u(2, 1, +)$ and $S_u(2, 1, -)$ are excited at $\Delta \nu = 0$~GHz, and $S_u(2, 1, +)$ is excited at $\Delta \nu = 4.14$~GHz as shown in Fig.~\ref{correlation-8.8mA}(a). We attribute this behavior to the splitting between the $S_u(2, 1, -)$ and $S_u(2, 1, +)$ modes, which seems to be about $\Delta \nu_O \approx 7.5$~GHz (see also Supplementary Section~\ref{ssec:orientation-splitting}): these two modes are excited alone when the $M_u(6, 1, \pm)$ modes aligns spectrally with them, and their superposition is created when $M_u(6, 1, \pm)$ is in between the two orientations of $S_u(2, 1)$. The same behavior is found when the $M_u(6, 1, \pm)$ modes come close to the $S_u(3, 1, \pm)$ modes [see Fig.~\ref{correlation-8.8mA}(b) and the magnification in Fig.~\ref{correlation-8.8mA}(c)]. We associate this scenario to the $M_u(6, 1, \pm)$ modes lasing simultaneously, which creates an azimuthally uniform intensity distribution that is able to excite both orientations of the slave laser modes equally well. 

However, spectral alignment alone does not guarantee strong synchronization. For instance, around $\Delta \nu \approx 115$~GHz only a low correlation around $9\%$ is observed despite the spectral matching between the $M_u(6, 1, \pm)$ modes and the $S_u(4, 1)$ mode. The third region of positive correlation shows a somewhat different scenario: at $\Delta \nu \approx 76$~GHz, the $M_u(5,1)$ mode of the master, which lases only in a single orientation, excites a superposition of the $S_u(2, 1, +)$ and $S_u(2, 1, -)$ modes. However, the correlation is low, probably because the $M_u(5,1)$ has less power than the $M_u(6, 1, \pm)$ modes. 

This second injection experiment confirms our earlier observation that spectral alignment between a strong mode of the master and a transverse mode of the slave laser is an important prerequisite for synchronization. While the transverse modes of the master and slave lasers that are coupled can be different, the spatial structure of the master laser mode can also play an important role: simultaneous lasing of the master laser mode in both orientations favors the emergence of three correlation peaks, in which different orientations of the same transverse mode of the slave laser are excited. Furthermore we find that a higher injection ratio does not necessarily lead to higher correlations,  which may be due to the master power being spread across more modes, while power concentration in a single mode seems to play a key role. 

Applying a low-pass filter improves the correlations to some extent, typically reaching peak values between 20\% and 30\% with cutoff frequencies around $0.5$ to $1$~GHz [see Fig.~\ref{correlation-8.8mA}(c)], but these values are below those found for the first case [see Fig.~\ref{correlation-6mA}]. We believe this is related to the type the master laser dynamics, which is chaotic with high frequency components up to $20$~GHz [see Fig.~\ref{Setup_free_running_V5}(c)], in contrast to the slower polarization-hopping dynamics in the first case. Whereas the slow polarization-hopping dynamics creates strong synchronization which is revealed by low-pass filtering, the chaotic dynamics in the second case relies on high-frequency components, so filtering does not help much. It seems that fast chaotic dynamics is harder to synchronize. The absence of inverse synchronization in the second case is explained by the absence of polarization-hopping dynamics to which we attribute the inverse synchronization in the first case. In summary, the second injection experiment confirms the importance of spectral alignment for synchronization [see also Supplementary Fig.~\ref{fig:2d-corr-map}]. However, we also find significant differences between the two cases, demonstrating the diversity of synchronization scenarios and their dependency on the dynamical and spatial properties of the master laser. 

\section*{Discussion}
Our experiment constitutes a lab-scale demonstration of synchronization between systems with complex spatio-temporal dynamics. By exploiting the intrinsic and controllable chaotic dynamics of broad-area VCSELs, it also constitutes the simplest and most practical architecture so far for multiple-user physical-layer secure communication schemes at high bit rates. Such systems would benefit from the highly complex and fast fluctuations of the laser dynamics and the spatial aspect of the dynamics involving a large number of laser modes. Moreover, the high-dimensional spatio-temporal dynamics of broad-area VCSELs could also be exploited in future photonic reservoir-computing architectures \cite{reservoir_2025}. 

Owing to the multimode nature and intrinsic complexity of the VCSELs, the dynamics of the coupled system is very rich, and we observe different scenarios including synchronization of ultrafast chaos, polarization-hopping dynamics, and inverse synchronization. The measured correlations are generally low of the order of $20\%$, which could be due to the very low transmission through the top mirror of the VCSELs, mismatch in the device structures and the intrinsic mode competition. However, correlations up to $90\%$ are observed for low-pass filtered polarization-hopping dynamics. Synchronization is typically observed when one of the dominant modes of the master laser is spectrally aligned with a mode of the slave laser, but does not require a careful matching of their spatial profiles. This flexibility and the multitude of possible synchronization scenarios gives hope that further improvements of the synchronization quality are possible. Still, the high correlation of the low-pass filtered polarization-hopping dynamics is promising for applications such as private key sharing \cite{annovazzi-lodi_private_2010, jiang_secure_2017} that typically require both high-dimensional chaos and good synchronization quality, but not very high speeds.  

Our work opens new perspectives concerning the synchronization of complex lasers and beyond. We show that synchronization of the temporal dynamics does not require or imply that other system properties like the spectrum and the spatial profiles become identical as well. Hence a more precise exploration and classification of synchronization scenarios for coupled spatio-temporal systems will be needed, with implications in other scientific fields. Beyond the temporal dynamics, further investigations of the spatial dimension appear promising, for example by using different cavity shapes \cite{Brejnak2021, Kim2023b, Alkhazragi2023} or shaping the injected beam. Tuning the birefringence or introducing controlled anisotropies could help to isolate polarization states and their associated spatial modes. By systematically varying the number and nature of the excited modes, one could assess how the modal complexity influences the robustness, speed, or nature of synchronization. Such control would allow for a more precise exploration of the transition from low-dimensional to high-dimensional chaotic synchronization in spatially extended laser systems. 


\newpage

\section*{Materials and Methods}

\subsection*{Correlation measurement}
The synchronization quality between the master and the slave lasers is calculated using the well-known correlation coefficient $C(\Delta t)$
\begin{equation}
    C(\Delta t) = \frac{\langle[P_M(t) - \langle P_M \rangle][P_S(t+\Delta t) - \langle P_S \rangle]\rangle}{\sqrt{\langle[P_M(t) - \langle P_M \rangle]^2\rangle \langle[P_S(t)- \langle P_S \rangle]^2 \rangle}}
\end{equation}
where $P_M(t)$ and $P_S(t)$ are the output powers (i.e., time traces) of the master and slave lasers, respectively, that are measured with the high-speed photodetectors. The time average is denoted by $\langle \rangle$, and $\Delta t$ is a time shift. The correlation is computed as function of $\Delta t$ for each detuning $\Delta \nu$. Figs.~\ref{correlation-6mA}(b) and \ref{correlation-8.8mA}(c), however, only show the correlation at the $\Delta t$ for which it is maximal. This time shift of $\Delta t \approx 3.78$~ns corresponds to the path-length difference between the master and slave signals, and the maximum of the correlation is indeed obtained at this delay, confirming that zero-lag synchronization is computed and observed. For detunings where the master and slave lasers are not synchronized, the correlation is theoretically expected to vanish. However, a persistent residual correlation remains observable in our system, fluctuating around $C = 3\%$ for the unfiltered data. This non-vanishing baseline arises from the reflection of the master laser from the top mirror of the slave laser. We verified that even with the slave laser switched off, a small fraction of the master signal is reflected back, producing a weak but measurable correlation. This weak correlation also appears at a time lag $\Delta t$ corresponding to the propagation delay along the signal pathways.

\subsection*{Filtering of time traces}
In addition to analyzing the correlation of the original measured time traces, we also investigated the effect of spectral filtering on the correlation.  A Butterworth low-pass filter was chosen due to its simplicity of implementation and its flat frequency response in the passband, which ensures that the preserved frequency components are not distorted by filtering artifacts. Different cutoff frequencies $f_c$ were tested, ranging from the full bandwidth of the measured signal (23~GHz) down to 10~MHz, allowing us to investigate how the correlation evolves across different frequency ranges. 

The transfer function of an $n$th-order Butterworth filter as function of RF-frequency $f$ is given by
\begin{equation} H(f) = \frac{1}{\sqrt{1 + (f/f_c)^{2n}}} 
\end{equation}
where $f_c$ is the cutoff frequency. It corresponds to the point where the transfer function drops to -3 dB, marking the boundary between the passband and the stopband. In our case, a filter of order $n = 2$ was used, providing a suitable trade-off between high-frequency attenuation and filter stability. We verified that the type and order of the low-pass filter (Butterworth, zero-phase, or linear-phase FIR) do not affect the synchronization coefficient, indicating that filtering does not distort the phase or delay in a way that impacts our results.\newline

\section*{Acknowledgements}
The Chair in Photonics is supported by Region Grand Est, GDI Simulation, Departement de la Moselle, European Regional Development Fund, CentraleSup\'elec, Fondation CentraleSup\'elec, and Eurometropole de Metz.

\section*{Contributions}
The experimental setup was jointly conceived by J.M., M.S. and S.B., and built and implemented by J.M., who also carried out the measurements and data analysis. S.B. and M.S. supervised the project. All authors discussed the results, and contributed to the writing of the manuscript.

\section*{Data availability}
Data underlying the results presented in this paper are not publicly available at this time but may be obtained from the authors upon reasonable request.

\section*{Conflict of interest}
The authors declare no conflicts of interest.


\bibliographystyle{naturemag}
\bibliography{apssamp.bib}


\onecolumngrid
\newpage

\begin{center}
\huge{Supplementary Information for "Synchronization of complex spatio-temporal dynamics with lasers"}\\
\phantom{adf} \\ 
\large{Jules Mercadier$^{*,1,2}$, Stefan Bittner$^{1,2}$ and Marc Sciamanna$^{*,1,2}$} \\ 
$^1$Université de Lorraine, CentraleSupélec, LMOPS F-57000 Metz, France\\
$^2$Chaire Photonique, LMOPS, CentraleSupélec, 57070 Metz, France \\
$^*$ Corresponding authors : jules.mercadier@centralesupelec.fr; marc.sciamanna@centralesupelec.fr

\end{center}
\twocolumngrid

\setcounter{page}{1}


\appendix
\renewcommand{\appendixname}{Supplementary Section}
\renewcommand{\thesection}{\arabic{section}} 
\renewcommand{\thesubsection}{\Alph{subsection}} 

\counterwithout{equation}{section} 

\renewcommand{\theequation}{S\arabic{equation}}
\renewcommand{\thefigure}{S\arabic{figure}}
\setcounter{figure}{0}
\setcounter{equation}{0}

\begin{figure*}[t]
    \centering
    \includegraphics[width=\linewidth]{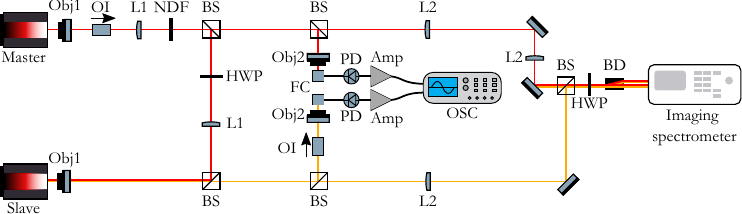}
    \caption{Experimental setup. Obj1: 40x objective (NA = 0.6); Obj2: 20x objective (NA = 0.5); OI: optical isolator; NDF: neutral density filter; BS: beam splitter; HWP: half-wave plate; FC: fiber coupler; PD: Photodetector; Amp: RF-Amplifier; OSC: oscilloscope; L1 (L2): lens with 200~mm (100~mm) focal length; BD: beam displacer.}
    \label{fig:SM_Setup}
\end{figure*}

\section{Experimental setup} \label{sec:setup}

The experimental setup is presented in Fig.~\ref{fig:SM_Setup} in more detail. The lasers under study are broad-area VCSELs (Frankfurt Laser Company FL85-F1-P1N-AC) with a circular output aperture of $15~\mu$m diameter, electrically pumped through a ring contact. The output beam of the master laser is collimated by a 40× microscope objective (Obj1, NA = 0.6). An optical isolator (OI) is placed immediately after the master laser to prevent any undesirable optical feedback that could influence its intrinsic dynamics. A beam splitter (reflexion: 70 / transmission: 30) directs part of the master laser signal to the detection arm (30\%), and part of the signal (70\%) is injected into the slave laser. In the same way, part of the slave laser signal is collimated and sent to the detection arm. 

The detection arm allows to measure the temporal dynamics, the optical spectra, and spatio-spectral images of the two lasers. Two fiber couplers with 20× objectives (Obj2, NA = 0.5) are employed to couple the beams into multimode graded-index fibers (Thorlabs M116L02, 50 $\mu$m core diameter) connected to high-speed photodetectors (Newport 1484-A-50, 22 GHz bandwidth). The RF-signals from the photodetectors are amplified by RF-amplifiers (SHF S126A) yielding a gain of 29 dB over a frequency range spanning from 80 kHz to 25 GHz, and measured by an oscilloscope (Tektronix DPO72340SX, 23 GHz bandwidth). Time traces were recorded with a sampling rate of 50 GS/s, using a record length of $10^6$ points (corresponding to a total acquisition time of $20~\mu s$). Because the detection chain is AC-coupled, the DC component of the optical signal is removed. The measured voltage therefore reflects only the temporal fluctuations of the laser’s optical power, excluding its constant (DC) offset. Before computing the correlation, the mean value of each time trace was subtracted. No additional detrending was applied. The normalization of the variance is inherently included in the definition of the correlation coefficient. Alternatively, the multimode fibers can be connected to an optical spectrum analyzer (OSA, Anritsu MS9740A, $30$~pm resolution) to examine the spectral composition of the laser emission. Optical isolators are used to prevent reflections from the fiber facets, and the entrance polarizers of the isolators are aligned to transmit only the $u$-polarization of the lasers (unless otherwise noted). 

By means of several lenses (L2, 100~mm focal length) and beam splitters, the output facets of the two BA-VCSELs are imaged onto the entrance slit of an imaging spectrometer (Princeton Instruments SpectraPro HRS-500), with a spectral resolution of 30 pm and 1800 g/mm grating. Due to the different beam paths and lens configurations used, the magnification is different for each laser. The slit of the spectrometer is fully open so that the camera in the focal plane of the spectrometer measures the spectrally-resolved near-field intensity distributions of the two VCSELs, that is, the intensity patterns of the different transverse modes are dispersed horizontally as function of their wavelength. The images of the two VCSELs are offset vertically, and a beam displacer (Thorlabs BD27, 2.7 mm beam separation) preceded by a half-wave plate is used to vertically separate the $u$- and $v$-polarizations of the slave laser as well. Thus the two polarizations of the slave laser and the $u$-polarization of the master laser are measured in parallel. \\ 
The signal of the master laser passes through the collimation objective (Obj1, NA = 0.6), a lens (L1, 200~mm focal length), another lens of the same type and finally the objective (Obj1) in front of the slaver laser. The lenses are aligned such that the output facet of the master laser is imaged onto the output facet of the slave laser with 1x magnification. Furthermore, additional cameras in the far-field planes of the lasers (not shown in Fig.~\ref{fig:SM_Setup}) are used to ensure that the injected beam is parallel to the output beam of the slave laser. The optical isolator after the master laser is used to select only its $u$-polarization for transmission to the slave laser, and a half-wave plate is used to align the polarization of the master signal with the $u$-polarization axis of the slave laser for parallel optical injection. A variable neutral density filter allows to change the power of the injected signal. The results presented in this article were obtained for minimal attenuation (maximal injection power). \\ 
As previously discussed, although the injection ratio is difficult to estimate precisely in practice, it remains very low and does not appear to directly affect the observed correlation levels. Another crucial factor is the optical alignment, which is essential to ensure optimal coupling between the two lasers. In addition to aligning the $u$-polarizations to allow for parallel injection, the angle at which the master beam enters the slave laser cavity had to be finely adjusted. To verify and optimize this alignment, both near-field and far-field profiles of the master and slave lasers were compared and carefully overlapped. This alignment proved to be crucial and technically challenging to implement.

\section{Properties of free-running lasers} \label{sec:free-running}

\subsection{Polarization and spectrum} \label{ssec:pol-spectrum}

\subsubsection{LI-curves and polarization-switching points}

\begin{figure}[tb]
    \centering
    \includegraphics[width=1\linewidth]{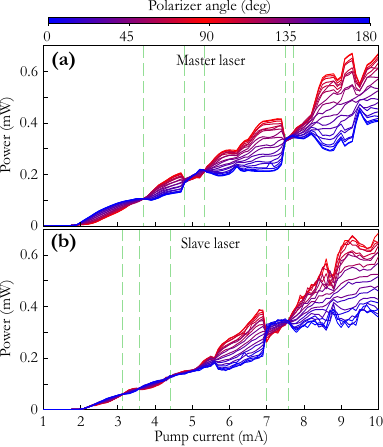}
    \caption{LI-curves of master (a) and slave laser (b) for 19 polarizer angles ranging from 0$^\circ$ to 180$^\circ$. The color corresponds to the polarizer angle, where $u$-polarization (at 90°) is shown in red and $v$-polarization (at 0$^\circ$ and 180$^\circ$) in blue. The dashed green lines indicate the polarization switching points (PSPs).}
    \label{fig:SM_LI_Curves}
\end{figure}

To identify the principal polarization axes of the BA-VCSELs used in this work, a polarizer in a motorized rotation stage was employed to measure the optical power transmitted through the polarizer as function of its orientation for both lasers. Using a power meter, measurements were taken every 10$^\circ$ over the full current range of 2 to 10~mA as shown in Fig.~\ref{fig:SM_LI_Curves}. The resulting profiles exhibit two distinct extrema in emitted power at 0$^\circ$ (or equivalently 180$^\circ$) and 90$^\circ$, revealing the two orthogonal polarization axes. Depending on the pump current and the temperature of the laser, the dominant polarization switches between these two orientations at specific PSPs, while the $u$- and $v$-polarization axes themselves remain fixed. In most cases, the 90$^\circ$ axis (referred to as $u$-polarization) dominates, while the 0$^\circ$ / 180$^\circ$ axis is referred to as $v$-polarization. See Ref. [S1] for more details. It should be noted that we define the polarizer angle with respect to the polarization axes of the VCSELs as a convenient reference frame to facilitate visualization and comparison. These axes generally do not coincide with the horizontal and vertical axes of the laboratory frame.

\subsubsection{Optical spectrum and birefringence}

\begin{figure}[tb]
    \centering
    \includegraphics[width=1\linewidth]{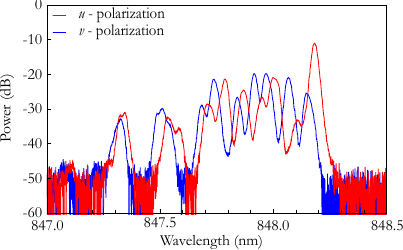}
    \caption{Optical spectra of the master laser at a pump current of $I_M = 6.5$~mA and a temperature of $T_M = 20^\circ$C, measured along the $u$- (red) and $v$-polarization (blue) axes.}
    \label{fig:SM_spec-birefringence}
\end{figure}

Measurements of the optical spectra provide important information on the properties of the lasers under study. First, the wavelengths of the individual modes are clearly identified, with the fundamental mode appearing at the longest wavelength (right side of the spectrum), and higher-order transverse modes at shorter wavelengths (as discussed in more detail in Ref. [S1]). Figure~\ref{fig:SM_spec-birefringence} shows the optical spectra of the master laser measured along the two dominant polarization axes, $u$ and $v$, for a fixed pump current of $I_M = 6.5$~mA. This measurement reveals that the optical power is distributed differently among the transverse modes depending on the polarization, and also highlights a spectral shift between the two polarization components. We assign the label "red" ("blue") to the $u$- ($v$-) polarization whose spectrum is at longer (shorter) wavelengths. 

This spectral separation is due to the intrinsic birefringence of the VCSEL which plays a critical role in polarization dynamics. We experimentally measured the birefringence frequency from the spectral separation between the same transverse modes in the two orthogonal polarizations. Although the birefringence may vary slightly depending on the mode, this analysis allowed us to estimate $\Delta \nu_b = 8.8$~GHz for the master laser and $\Delta \nu_b = 8.4$~GHz for the slave laser. We explicitly chose two VCSELs with very similar birefringence and close central wavelength value. 

\subsubsection{Polarization-hopping dynamics}

\begin{figure}[tb]
    \centering
    \includegraphics[width=1\linewidth]{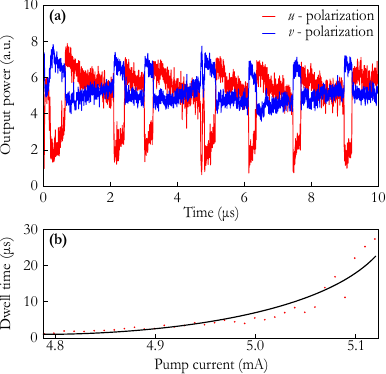}
    \caption{(a) Time traces of the master laser at $I_M$= 4.95~mA, low-pass filtered at a cutoff of 0.1~GHz, for the two orthogonal polarization axes $u$ (red) and $v$ (blue). (b) Mean dwell time of the time traces in one polarization state as function of the pump current.}
    \label{fig:SM_hopping}
\end{figure}

The VCSELs under study exhibit several polarization switching points (PSPs) at which the dominant polarization changes (see Fig.~\ref{fig:SM_LI_Curves}), such as in the case of the master laser at 3.6~mA, as well as points at which the power in the two polarization states becomes almost equal without switching, such as the master laser at 6.2~mA. These points are sometimes accompanied by a dynamical regime known as polarization hopping. As illustrated in Fig.~\ref{fig:SM_hopping}(a), this regime is characterized by intensity fluctuations in one polarization ($u$), with an anti-correlated response in the orthogonal polarization ($v$). Due to the relatively long timescales of this dynamics, the associated intensity fluctuations contribute significantly to the low-frequency components of the RF spectrum, predominantly below $500$~MHz. 

This behavior has been reported in various VCSEL systems [S2-S4], and the spontaneous polarization switching can be driven by noise or by deterministic nonlinear dynamics. The temporal behavior within this regime is characterized by the dwell time, defined as the average duration the system remains in one polarization state before switching. Figure~\ref{fig:SM_hopping}(b) presents the measured dwell time as a function of pump current, showing an exponential increase with increasing current. This trend is characteristic of deterministic polarization chaos [S4]. 

\subsubsection{Orientation of transverse modes} \label{ssec:orientation-splitting}

\begin{figure}[tb]
    \centering
    \includegraphics[width=0.9\linewidth]{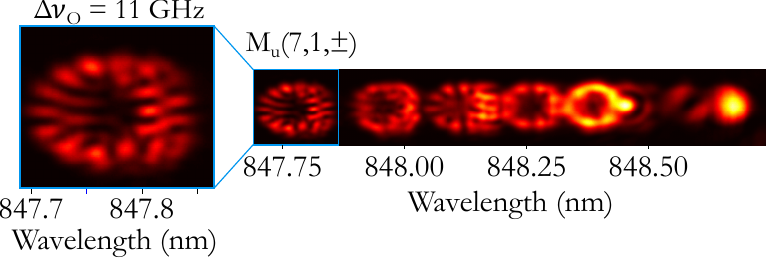}
    \caption{Spatio-spectral image of the free-running master laser in $u-$polarization for $I_M$= 9.8 mA with the modes $M_u(7,1,+)$ and $M_u(7,1,-)$ highlighted in the inset in blue. They are offset laterally due othe the non-vanishing orientation splitting $\Delta \nu_O \approx 11$~GHz.}
    \label{fig:SM_splitting_transverse}
\end{figure}

Due to the circular geometry of the VCSEL cavity, all transverse modes $(m, n)$ with $m > 0$ exist in two different orientations, designated by $+$ ($-$) if they are (anti-) symmetric with respect to a chosen axis. In an ideally symmetric cavity, the modes with $+$ and $-$ orientations are degenerate. 

Figure~\ref{Setup_free_running_V5}(b$_2$) presents polarization-resolved spatio-spectral measurements of the master laser under free-running conditions for $I_M$= 6.5 mA. Several key observation can be made: first, we typically observe modes with the same transverse pattern in both polarization states, however, there is a clear spectral shift between them due to the birefringence as discussed above. Second, the intensities in the two polarizations are generally not the same (see Fig.~\ref{fig:SM_spec-birefringence}). Third, for a given $(m, n)$, the $u$- and $v$-polarizations usually exhibit different transverse orientations, with a $-$ symmetry in $u$-polarization corresponding to a $+$ symmetry in $v$-polarization, which is attributed to gain competition [S5,S6]. In short, usually each transverse mode lases in both polarizations but with different spatial orientations and different intensities. However, in some cases, such as mode $M_u(6,1)$ in Fig.~\ref{correlation-8.8mA}, the measured spatial pattern shows that both orientations lase in the same polarization state (under free-running conditions). 

Figure~\ref{fig:SM_splitting_transverse} shows the spatio-spectral image of the master laser at 9.8~mA. In this example, the mode $(7,1)$ lases in both orientations in $u$-polarization, however, we see that the $M_u(7,1,+)$ and the $M_u(7,1,-)$ mode are not degenerate as expected for a perfectly symmetric system, but spectrally shifted. We designate this frequency separation between the $+$ and $-$ orientations of a given transverse mode within the same polarization as $\Delta \nu_O$. 
This splitting is usually below the resolution of standard optical spectrum analyzers or imaging spectrometers, making it difficult to measure directly. In the case of Fig.~\ref{fig:SM_splitting_transverse}, it is quite large with $\Delta \nu_O \approx 11$~GHz. In other instances, like the mode $M_u(6,1)$ in Fig.~\ref{correlation-8.8mA}, it is close to $0$. 

Moreover, the progressive excitation of different orientations under optical injection, such as the modes $S_u(2,1)$ and $S_u(3,1)$ previously presented in Fig.~\ref{correlation-8.8mA}, enables an indirect estimation of $\Delta \nu_O$ around 7.64~GHz and 8~GHz, respectively, by considering the secondary correlation peaks observed near the principal one. Overall, the degeneracy splitting $\Delta \nu_O$ is strongly mode-dependent and may range from nearly 0 up to at least 11~GHz, highlighting the complexity of polarization and orientation resolved transverse mode spectra in BA-VCSELs. 

\subsection{Comparison of master and slave laser} \label{ssec:laser-comp} 

Among a batch of five nominally identical BA-VCSELs from the same manufacturer, a detailed characterization was performed on each device to identify the most suitable pair for the injection experiment. The selection was guided by two key criteria: similarity in birefringence frequency $\Delta \nu_b$ and in wavelength of the lasing modes. As the laser dynamics are strongly influenced by birefringence, we selected two devices exhibiting closely matched values, $\Delta \nu_b = 8.8$~GHz for the master laser and $\Delta \nu_b = 8.4$~GHz for the slave. Spectral alignment was also critical, since frequency detuning strongly affects the injection dynamics. Both lasers exhibited comparable emission wavelengths over the considered current range ($2$–$10$~mA), ranging from $847.6$ to $848.6$~nm at a common temperature of 20\textdegree{}C. 

\begin{figure}[tb]
    \centering
    \includegraphics[width=1\linewidth]{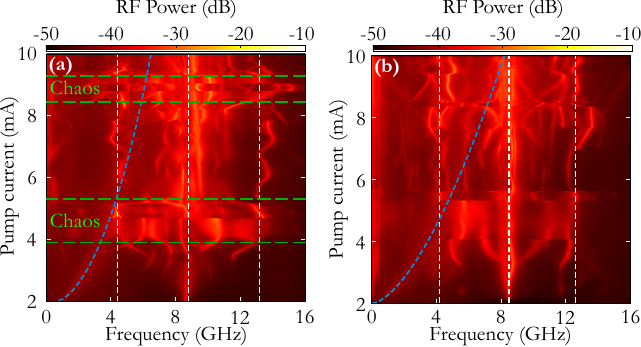}
    \caption{RF-spectra of the master (a) and slave laser (b) as a function of the pump current from 2 to 10 mA. The RF-spectra were measured for a polarizer angle of $45^\circ$. The RF-spectra are smoothened by a sliding average over 10 points (0.5 MHz). The birefringence frequencies $\Delta \nu_b$ as well as 0.5$\Delta \nu_b$ and 1.5$\Delta \nu_b$ are indicated with white dashed lines, and the relaxation oscillation frequency with blue dashed lines. The green dashed lines in (a) delimit the current regimes in which the master lasers exhibits high-frequency chaotic dynamics.}
    \label{fig:SM_RF_Spectra}
\end{figure}

Prior to the injection experiment, the intrinsic free-running behavior of each laser was analyzed. Figure~\ref{fig:SM_LI_Curves} presents polarization-resolved LI curves for both devices. While both exhibit well-defined orthogonal polarization axes, their PSPs differs significantly, even though the $u$-polarization (red) generally dominates, which is the polarization axis we chose for injection. Further insights were obtained from the analysis of the RF-spectra shown in Fig.~\ref{fig:SM_RF_Spectra}, which displays the evolution of RF-spectra as function of the pump current measured along the 45\textdegree{} polarization axis, that is, $u$- and $v$-polarization are measured with equal efficiency on the photodetector. Consistent with previous studies [S7], the dominant frequency components appear around the birefringence frequency $\Delta \nu_b$, as well as its subharmonic and harmonic (0.5$\Delta \nu_b$ and 1.5$\Delta \nu_b$, white dashed lines). Relaxation oscillations are also visible (blue dashed lines). The bifurcations, often coinciding with PSPs, are clearly reflected in changes of the spectral content. Some bifurcations mark the begin of a regime of high-frequency chaos, delimited by the green dashed lines in Fig.~\ref{fig:SM_RF_Spectra}(a) for the master laser. The chaotic regions were identified using the noise titration method as well as the Grassberger-Procaccia algorithm [S7]. 

These analyses show that although the two selected VCSELs share key parameters such as birefringence and emission wavelength, their intrinsic dynamics are far from being stationary and furthermore differ significantly. This makes polarization-resolved injection experiments between them a challenging task. Once the pair of VCSELs was identified, polarization alignment in the injection path was ensured using a half-wave plate (see Fig.~\ref{fig:SM_Setup}). Due to intrinsic birefringence and structural anisotropies, the $u$- and $v$-polarization axes of each VCSEL differ: for the master, $u = 160^\circ$ and $v = 70^\circ$; for the slave, $u = 150^\circ$ and $v = 60^\circ$ (note that these angles are given with respect to the horizontal axis of the laboratory frame, in contrast to the convention established in section~\ref{ssec:pol-spectrum}). Combined with the 45\textdegree{} rotation introduced by the optical isolator and further rotations from subsequent reflections, proper alignment of the injected field was achieved by tuning the HWP to ensure that the $u$-polarization of the master was injected into the $u$-axis of the slave to achieve polarization matching. 

Furthermore, the estimated relaxation oscillation frequencies range from 0.7 to 6.7 GHz for the master laser and from 0.6 to 7.9 GHz for the slave laser as a function of pump current, but no significant influence of their matching on the synchronization quality is observed in our experiments.

\subsection{Dependency on laser parameters} \label{ssec:param-depend}
The emission properties of the BA-VCSELs exhibit a strong dependency on both pump current and temperature. As expected for semiconductor lasers, increasing the pump current induces a red-shift of the emission wavelength due to Joule heating. In addition, the transverse mode structure evolves with the current, often accompanied by polarization-dependent redistribution of intensity among the modes. For instance, higher-order modes emerge at increased current levels, with distinct spectral positions and polarization preferences. Thermal effects further shift the emission spectrum. A linear red-shift is observed as a function of temperature and pump current, with typical tuning coefficients of approximately 0.05~nm/\textdegree{}C and 0.136~nm/mA. These shifts impact the spatial overlap  and the spectral alignment between the two VCSELs. Accurate current and temperature control is therefore critical to maintain spectral alignment, especially in weak injection regimes.

\section{Injection experiments}

\subsection{Injection ratio} \label{ssec:injection-ratio}

The injection ratio is a fundamental parameter for coupling two laser systems. In this study, it is quantified as the injection strength $\kappa_{\mathrm{inj}}$, calculated as the power from the master laser in the $u$-polarization at the input of the slave laser divided by the total output power of the slave laser $P_{\mathrm{S,fr}}$, 
\begin{equation}
\kappa_{\mathrm{inj}} = \frac{P_{\mathrm{inj}}}{P_{\mathrm{S,fr}}} \, 
\end{equation}
The measured injection power accounts for optical losses due to the isolator, beam splitter, and other optical elements. Furthermore, the injection strength can be reduced via a variable neutral density filter (see Fig.~\ref{fig:SM_Setup}), but it was set to minimal attenuation in the experiments presented here. $P_{\mathrm{S,fr}}$ is the free-running total output power of the slave laser, measured directly at the output, and includes both polarization components.

\begin{figure}[tb]
    \centering
    \includegraphics[width=0.9\linewidth]{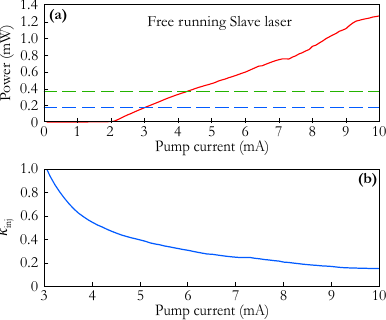}
    \caption{(a) Total output power $P_{\mathrm{S,fr}}$ of the slave laser in free-running conditions as a function of the pump current. The green (blue) dashed lines represent the injected power from the master laser for $I_M = 8.8\,\mathrm{mA}$ ($I_M = 6.17$~mA), respectively. (b) Corresponding injection strength $\kappa_{\mathrm{inj}}$ between the master and slave lasers for $I_M = 6.17$~mA, plotted as a function of the slave laser's pump current $I_S$.}
    \label{fig:SM_LI_Slave_FR}
\end{figure}

In the first case of injection studied in the paper, the master laser operates at fixed parameters ($I_M = 6.17\,\mathrm{mA}$) and delivers an input power of approximately $P_{\mathrm{inj}} = 0.18$~mW at the entrance of the slave laser, indicated by the blue dashed line in Fig.~\ref{fig:SM_LI_Slave_FR}(a). The output power of the slave $P_{\mathrm{S,fr}}$ varies nearly linearly with pump current as shown in Fig.~\ref{fig:SM_LI_Slave_FR}, and this variation determines the resulting injection strength $\kappa_{\mathrm{inj}}$, plotted in Fig.~\ref{fig:SM_LI_Slave_FR}(b). It is important to note that these values do not take into account reflections at the slave input facet or other factors affecting the coupling efficiency; they only reflect the available optical power at that point. 

As the slave pump current $I_S$ increases, its output power increases, leading to a decrease in $\kappa_{\mathrm{inj}}$. For example, for the correlation peaks labeled $a_1$, $a_2$, $a_3$, and the final peak shown in Fig.~\ref{correlation-6mA}, corresponding to slave currents of approximately $6.0$, $7.5$, $8.5$, and $9.5$~mA, respectively, the injection strengths are estimated around 0.30, 0.22, 0.17, and 0.14. This represents a reduction by a factor of almost two of the injection strength between the first and last peak. However, the correlation values are roughly the same in all four cases, suggesting that moderate variation in injection strength does not significantly impact correlations in this regime. 
\\ 

In the second case of injection we study, at $I_M = 8.8$~mA and $I_S = 4.8$~mA, the injected power reaching the slave is approximately $0.37$~mW [green dashed line in Fig.~\ref{fig:SM_LI_Slave_FR}(a)], while the total output power of the slave laser is $0.44$~mW, corresponding to an injection strength of $\kappa_{\mathrm{inj}} = 0.84$. 

However, it is important to note that the values of $\kappa_{\mathrm{inj}}$ presented here should be interpreted with caution. The measured master laser power corresponds to the value at the front of the slave laser. It is not possible to directly determine the portion that actually couples into the slave cavity. A rough estimate can nevertheless be made. In fact, assuming a typical front-facet reflectivity of $> 90\%$, only a fraction of less than $10\%$ of the master power effectively enters the cavity. Similarly, the total output power of the slave laser used here does not reflect the real intracavity power, and is therefore underestimated. So while the injection strength is rather weak in our experiment, it allows us to rule out trivial synchronization effects driven purely by high-power injection. 

\subsection{Inverse synchronization}

\begin{figure}[tb]
    \centering
    \includegraphics[width=1\linewidth]{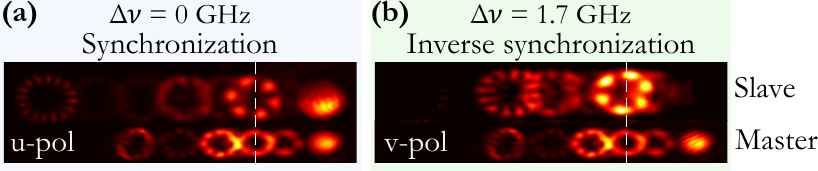}
    \caption{Spatio-spectral image of the master laser in $u-$polarization for $I_M$= 6.17 mA (bottom), with the spatio-spectral image of the slave laser at $\Delta \nu = 0$ in $u-$ polarization (a) and at $\Delta \nu = 1.7$~GHz in $v-$polarization (b).}
    \label{fig:spatio-spec-inv-synch}
\end{figure}

The inverse synchronization phenomenon observed in the first injection case has been attributed to polarization-related dynamics. Figure~\ref{fig:spatio-spec-inv-synch} considers the first synchronization region and shows in panel (a) the case of $\Delta \nu = 0$, where the $S_u(3,1)$ mode is spectrally aligned with the $M_u(3,1)$ mode. As a result, its intensity increases significantly, while neighboring modes are suppressed. This spectral alignment illustrates the strong mode interaction when their frequencies coincide. Figure~\ref{fig:spatio-spec-inv-synch}(b) shows a similar effect, but in the orthogonal polarization of the slave laser ($v$-axis) at $\Delta \nu = 1.7$~GHz: the $S_v(3,1)$ mode aligns with the $M_u(3,1)$ mode, leading to an intensity increase of $S_v(3,1)$ at the expense of adjacent modes. This alignment coincides with the occurrence of the negative correlation peak, suggesting anti-correlation between the $u$-polarized emission of the master and slave laser appears because the $v$-polarized emission of the slave is correlated with the master signal. This example illustrates that the excitation of a slave laser mode in the $v$-polarization systematically coincides with the occurrence of the inverse synchronization peaks.

\subsection{Injection with variation of slave current and temperature}

\begin{figure}[tb]
    \centering
    \includegraphics[width=1\linewidth]{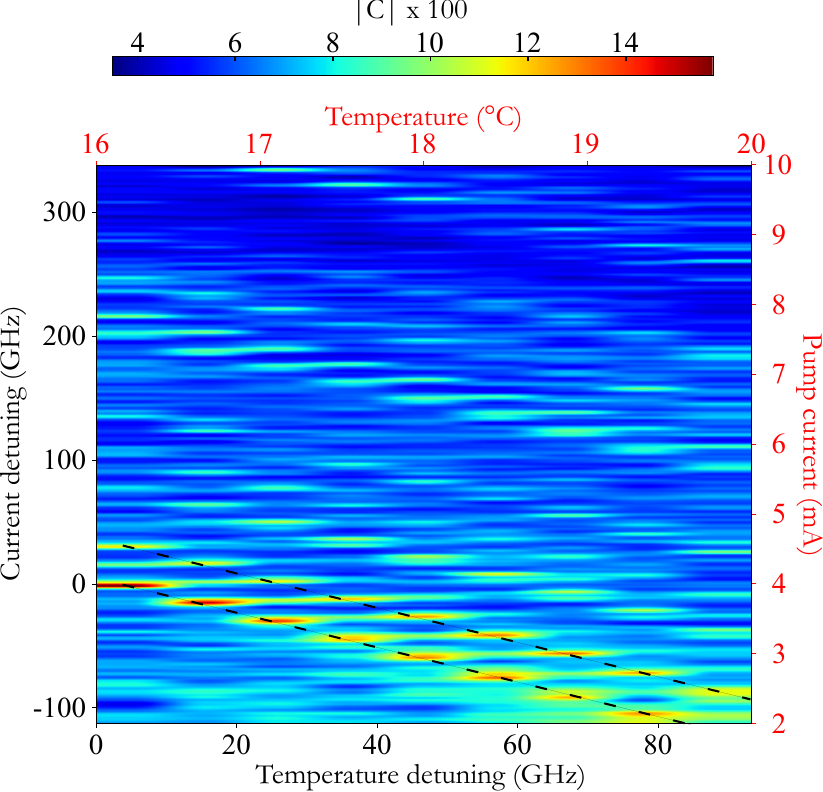}
    \caption{Correlation map between the master laser operated at $T_M = 19^\circ$C and $I_M = 9$~mA in the $u$-polarization, and the slave laser as a function of its temperature (horizontal axis) and injection current (vertical axis). The color scale represents the absolute value of correlation ($|C| \times 100$) of the unfiltered time traces. The black dashed lines follow the two principal series of correlation peaks.}
    \label{fig:2d-corr-map}
\end{figure}

Here, we present the evolution of the correlation between the master and slave lasers as a function of several experimentally scanned parameters. The master laser is held fixed at $I_M = 9$~mA and $T_M = 19^\circ$C, while the slave laser is scanned in both temperature ($0.5^\circ$ step) and injection current ($0.01$ mA step). The time traces are measured as function of the pump current, following the same procedure used in the first injection case discussed earlier, while keeping the slave temperature fixed. Then the slave temperature is increased followed by another sweep of the current, covering a range of 16$^\circ$C to 20$^\circ$C. 

Figure~\ref{fig:2d-corr-map} displays the absolute value of the correlation of the unfiltered time traces as function of slave temperature and current. The temperature and current shifts are converted to the corresponding detuning, and the total detuning $\Delta \nu$ is the sum of the current and temperature detunings. The map reveals several series of correlation peaks, with the first and strongest one emerging at $\Delta \nu = 0$~GHz. These correlation peak series appear as diagonals, reflecting conditions of spectral alignment between specific transverse modes of the master and slave lasers. The key implication of these diagonal series is that spectral alignment is a necessary condition for correlation, and can be achieved with both current and temperature variation. However, we also observe that the correlation varies along these diagonals, suggesting that other experimental parameters also influence the quality of synchronization.

\subsection{Enhancement of Slave laser mode}
\begin{figure}[tb]
    \centering
    \includegraphics[width=1\linewidth]{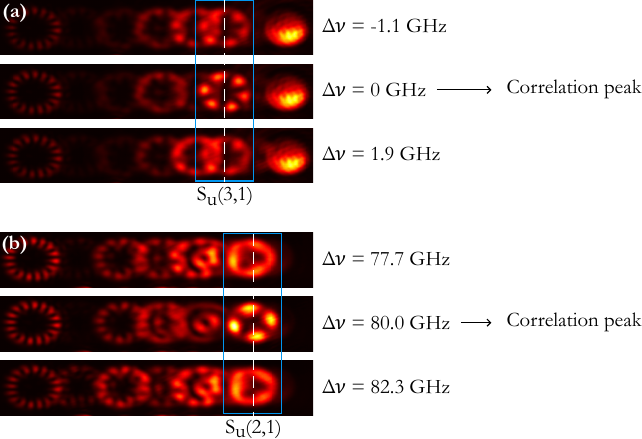}
    \caption{Spatio-spectral images of the injected slave laser in \( u \)-polarization for different values of the detuning $\Delta \nu$ for the first injection case with \( I_M = 6.17\,\mathrm{mA} \). (a) At \( \Delta\nu = 0\,\mathrm{GHz} \) we find the spectral alignment of the slave mode \( S_u(3,1) \) with the master mode \( M_u(3,1) \), (b) and at \( \Delta\nu = 80\,\mathrm{GHz} \) the alignment of \( S_u(2,1) \) with \( M_u(0,1) \). The spectrally aligned slave modes are highlighted by the blue rectangles.}
    \label{fig:SM_Slave_NF_Excited}
\end{figure}

The synchronization regions observed in the first injection case (\( I_M = 6.17\,\mathrm{mA} \)) are closely associated with spectral alignment between a strong transverse mode of the master laser and a transverse mode of the slave laser as function of the frequency detuning \( \Delta\nu \), see Fig.~\ref{correlation-6mA}. Several distinct correlation peaks were identified, notably at \( \Delta\nu = 0\,\mathrm{GHz} \) and \( \Delta\nu = 80\,\mathrm{GHz} \), each corresponding to the alignment of specific lasing modes. When the slave mode \( S_u(3,1) \) becomes spectrally aligned with the master mode \( M_u(3,1) \) at \( \Delta\nu = 0\,\mathrm{GHz} \), the intensity of \( S_u(3,1) \) increases significantly while that of neighboring modes decreases due to the injection as shown in Fig.~\ref{fig:SM_Slave_NF_Excited}(a). This selective excitation of \( S_u(3,1) \) is accompanied by a clear correlation peak, confirming the onset of synchronization. In contrast, for slightly detuned values such as \( \Delta\nu = -1.1\,\mathrm{GHz} \) or \( \Delta\nu = 1.9\,\mathrm{GHz} \), the spectral alignment is lost and the mode \( S_u(3,1) \) is less intense, which goes along with a decrease in correlation. 

A similar behavior is observed near \( \Delta\nu \approx 80\,\mathrm{GHz} \), where spectral alignment occurs between the slave mode \( S_u(2,1) \) and the master mode \( M_u(0,1) \). 
The intensity of the slave mode increases like in the first example, indicating its enhanced excitation due to spectrally aligned injection, and this results in a secondary synchronization peak. These observations demonstrate that the enhancement of a certain transverse mode of the slave laser, driven by spectral alignment with a strong mode of the master, plays a central role in enabling synchronization, but matching the spatial profiles of the involved master and slave modes is not required. 

\subsection{Correlation evolution with low- and high-pass filters}

\begin{figure}
    \centering
    \includegraphics[width=1\linewidth]{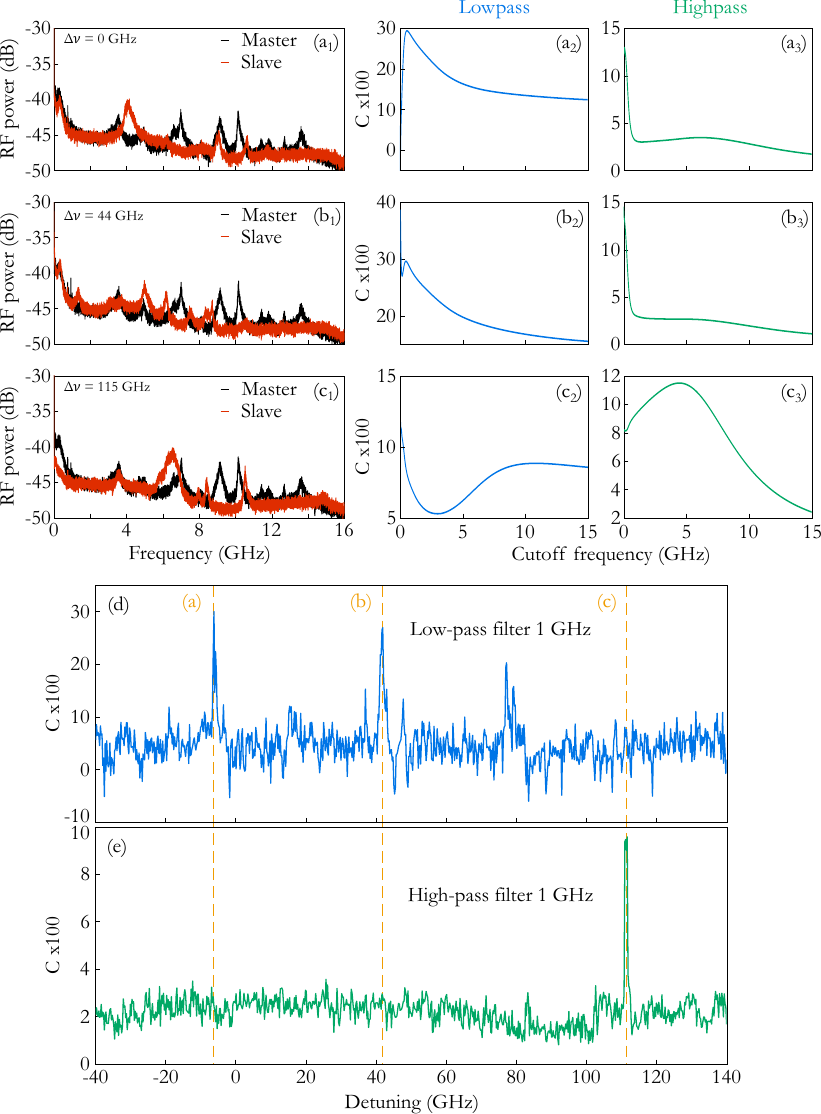}
    \caption{(a,b,c$_1$) RF-spectra of the master (black) and slave (red) lasers. Evolution of the correlation as a function of cutoff frequency for low-pass (a,b,c$_2$) and high-pass (a,b,c$_3$) filtering for three detuning values: (a) \( \Delta\nu = 0\,\mathrm{GHz} \), (b) \( \Delta\nu = 44\,\mathrm{GHz} \), and (c) \( \Delta\nu = 115\,\mathrm{GHz} \). (d)  Correlation as a function of detuning using a low-pass filter at 1 GHz cutoff and (e) a high-pass filter at 1 GHz.}
    \label{fig:SM_highpass}
\end{figure}

The preferential synchronization of low-frequency components appears to depend on the system and on the specific synchronization region. As shown in Fig.~\ref{correlation-6mA}(d) of the main article, the correlation value increases as the high-frequency components are progressively removed, thereby revealing one of the characteristic timescales that favors higher correlation value.

Here, we present the evolution of the correlation values between the master and slave lasers for three correlation peaks, associated with the second injection experiment shown in Fig.~\ref{correlation-8.8mA}. We determine how the correlation evolves when applying a low- or high-pass filter, with the corresponding results displayed in Fig.~\ref{fig:SM_highpass}.

These measurements of the evolution of the correlation value under low-pass and high-pass filtering show that, for the first two correlation peaks (a) and (b), the correlation increases as the high-frequency components are removed (Fig.~\ref{fig:SM_highpass} ($a_2$), ($b_2$)). The region corresponding to the third peak (c), at a detuning of 115 GHz, exhibits a lower synchronization level, associated with faster timescales. This ability to easily synchronize low-frequency components is consistent with the literature [S8-S11] and with the majority of our experiments.

\section*{References}
\noindent 
[S1] Bittner, S. \& Sciamanna, M. Complex nonlinear dynamics of polarization and transverse modes in a broad-area vcsel. \textit{APL Photonics} \textbf{7}, 126108 (2022).\\[2pt]
[S2]  Martin-Regalado, J. \textit{et al}. Polarization properties of vertical-cavity surface-emitting lasers. \textit{IEEE Journal of Quantum Electronics} \textbf{33}, 765–783 (1997).\\[2pt]
[S3] Willemsen, M. B. \textit{et al}. Polarization Switching of a Vertical-Cavity Semiconductor Laser as a Kramers Hopping Problem.\textit{Physical Review Letters} \textbf{82}, 4815-4818 (1999).\\[2pt]
[S4] Virte, M. \textit{et al}. Deterministic polarization chaos from a laser diode. \textit{Nature Photonics} \textbf{7}, 60–65 (2013). \\[2pt]
[S5] Debernardi, P. \textit{et al}. Influence of anisotropies on transverse modes in oxide-confined vcsels. \textit{IEEE Journal of Quantum Electronics} \textbf{38}, 73–84 (2002).\\[2pt]
[S6] Mart\`in-Regalado, J., Balle, S. \& Miguel, M. S. Polarization and transverse-mode dynamics of gain-guided vertical-cavity surface-emitting lasers. \textit{Optics Letters} \textbf{22}, 460–462 (1997).\\[2pt]
[S7] Mercadier, J. \textit{et al}. Chaos from a free-running broad-area VCSEL. \textit{Optics Letters} \textbf{50}, 796–799 (2025).\\[2pt]
[S8] Virte, M., Sciamanna, M. \& Panajotov, K. Synchronization of polarization chaos from a free-running VCSEL. \textit{Optics Letters} \textbf{41}, 4492-4495 (2016).\\[2pt]
[S9] Boccaletti, S. \textit{et al.} The synchronization of chaotic systems. \textit{Physics Reports} \textbf{41}, 1-2 (2002).\\[2pt]
[S10] Mercadier, J. \textit{et al.} Optical chaos synchronization in a cascaded injection experiment. \textit{Optics Letters} \textbf{49}, 2613-2616 (2024).\\[2pt]
[S11] Hramov, A. E. and Koronovskii, A. A. Time scale synchronization of chaotic oscillators. \textit{Physica D} \textbf{206}, 3-4 (2005).\\[2pt]


\end{document}